\documentclass[epj]{svjour}
\usepackage{epsfig} 
\newcommand{\PO}{I\!\!P}
\newcommand{\xpom}{x_{\PO}}
\newcommand{\ypom}{y_{\PO}}
\newcommand{\pbinv}{\mbox{${\rm pb^{-1}}$}} 
\def\Journal#1#2#3#4{{#1} {\bf #2} (#3) #4}

\def\PLB{{\em Phys. Lett.} {\bf B}}

\def\PRD{{\em Phys. Rev.} {\bf D}}

\def\EPJ{{\em Eur. Phys. J.} {\bf C}}
\begin{document}
\title{Studies of DVCS and photoproduction of photons at high $t$ with the H1
Detector}
\author{Laurent Favart
\thanks{\emph{supported by the 
{\it FNRS} - Belgium}}%
}                     
%
%
\institute{I.I.H.E., Universit\'e Libre de Bruxelles, Belgium}
\date{Contribution to the proceedings of the EPS03 conference}
%
\abstract{The diffractive scattering of photon off proton 
$\gamma^{(*)} p \rightarrow \gamma Y$ 
(where $Y$ denotes the proton dissociative system or the elastically 
scattered proton)
is measured in two different kinematic regimes at high energy with the H1
detector at HERA. 
i) the process is studied for the first time at large transverse photon
momentum $p_t^{\gamma} > 2$ GeV in photoproduction regime ($Q^2 < 0.01$ GeV$^2$)
and photon-proton c.m.s. energy $175 < W < 247 \ {\rm GeV}$.
The cross section is measured differentially in $t$ and in $\xpom$. 
The measurement is compared to a LL BFKL prediction of QCD.
ii) in electroproduction, the Deeply Virtual Compton Scattering (DVCS)
$\gamma^* p \rightarrow \gamma p$ cross section has been measured
more precisely and in an extended kinematic domain:
at photon virtualities $4<Q^2<80$ GeV$^2$ and $30<W<140$ GeV.
The measurement is compared to NLO QCD calculations 
using the DGLAP approximation
and to Colour Dipole model predictions.
} 
\maketitle
\section{Introduction}
\label{intro}
The hard diffractive scattering of a photon off a proton:
$\gamma^{(*)} p \rightarrow \gamma Y$ 
(where $Y$ denotes the proton dissociative system or the elastically
scattered proton)
is an important process for the understanding of high energy diffraction.
Its study allows to investigate the applicability and the relevance of
perturbative Quantum Chromo Dynamics (QCD) and provides unique
information on the proton structure. It gives access 
to new class of parton distribution functions,
the Generalised Parton Distributions (GPD) 
that can be interpreted as parton correlation functions in the proton.
Compared to vector meson production, the photon scattering 
is theoretically simpler because the 
theoretical predictions avoid
uncertainties due to unknown meson wave functions.

In the present report, the process is studied 
in two different kinematic regimes.\\
i) the photoproduction of photons at high $t$ (also called high $t$ Virtual
Compton Scattering)\\
ii) the Deeply Virtual Compton Scattering (DVCS).\\
In the first case, the hard scale needed for perturbative QCD 
predictions is provided
by the large $t$ value, where $t$ is the squared 4-momentum transfer at
the proton vertex, in the second, by the large $Q^2$ value. In the first
case due to large $t$ values, the cross section is largely dominated 
by the proton dissociation, while in the second, the measurement is
performed in the elastic case ($Y=p$).
The measurements can be compared to QCD predictions, respectively,
in the two asymptotic
limits BFKL and DGLAP. 

\section{Photoproduction of photons at high $t$.}
\label{sec:high_t}
The data for this analysis were collected with the H1 detector during
the 1999-2000 running period, when HERA collided 27.6 {\rm GeV}
positrons with 920 {\rm GeV} protons. An integrated luminosity of 47.6
${\rm pb}^{-1}$ is used.

\noindent
Photoproduction events were selected by
detecting the scattered positron in the electron tagger of the
luminosity system restricting the virtuality of the photon to
$Q^{2} < 0.01 $ GeV$^{2}$ and the photon-proton centre of mass energy 
to $175 < W < 247 \ {\rm GeV}$.
 
\noindent
Photons with an energy of at least 8 GeV were identified in the 
electromagnetic part of the SPACAL calorimeter 
(covering the polar region $153^{\circ} < \theta < 177.8^{\circ}$
\footnote{$\theta$ is measured relative to the
  outgoing proton beam direction.}).
The hadronic final state, if visible, is measured in 
a liquid argon (LAr) calorimeter covering the range in polar angle
$4^{\circ} < \theta < 153^{\circ}$.  
Diffractive events were selected by requiring that $\ypom < 0.018$,
where
\begin{equation}
\ypom = \frac{p.(q-q')}{q.p} \simeq \frac{\sum_Y(E-P_z)}{2E_{\gamma}}.
\end{equation}
$p$ and $q$ being the 4-vectors of the incoming proton and photon
respectively and $q'$ the 4-vector of the scattered photon. The
quantity is calculated experimentally by summing the $E-P_z$ of all
hadronic final state objects in the event (i.e. all measured particles
except the scattered electron and high $p_t$ photon), and dividing by
twice the incoming photon energy $E_\gamma$. 
As $\ypom \simeq e^{-\Delta
  \eta}$, this cut ensures that there is a large rapidity gap between
the photon and the proton dissociative system \footnote{To ensure
  efficient background rejection, a minimum rapidity gap of $\Delta
  \eta = 2$ is required between the photon candidate and the edge of
  the proton dissociative system.}. However, no proton dissociative system
is required to be seen in the detector.
 
\noindent
In addition to the kinematic variables defined above, the variable
$\xpom$ is defined as
\begin{equation}
\label{xpom}
\xpom = \frac{q.(p-Y)}{q.p} \simeq \frac{(E+P_z)_\gamma}{2E_{p}},
\end{equation}
where $Y$ is the 4-vector of the proton dissociative system,
$(E+P_z)_\gamma$ is the $E+P_z$ of the photon candidate and $E_{p}$ is
the energy of the incident proton.

\begin{figure}
 \epsfig{file=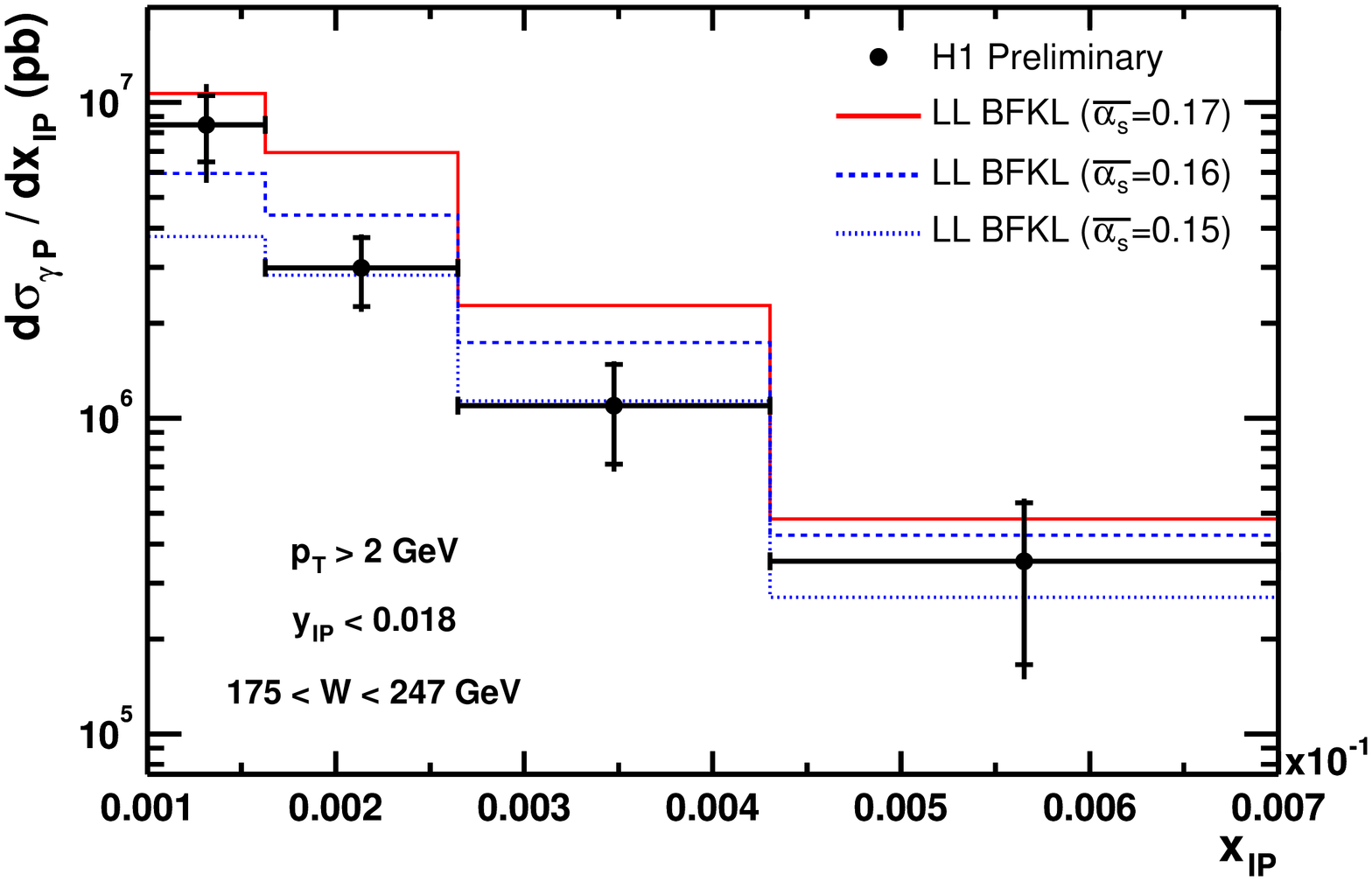,angle=270, width=.48\textwidth}
 \\[1.5ex]
 \epsfig{file=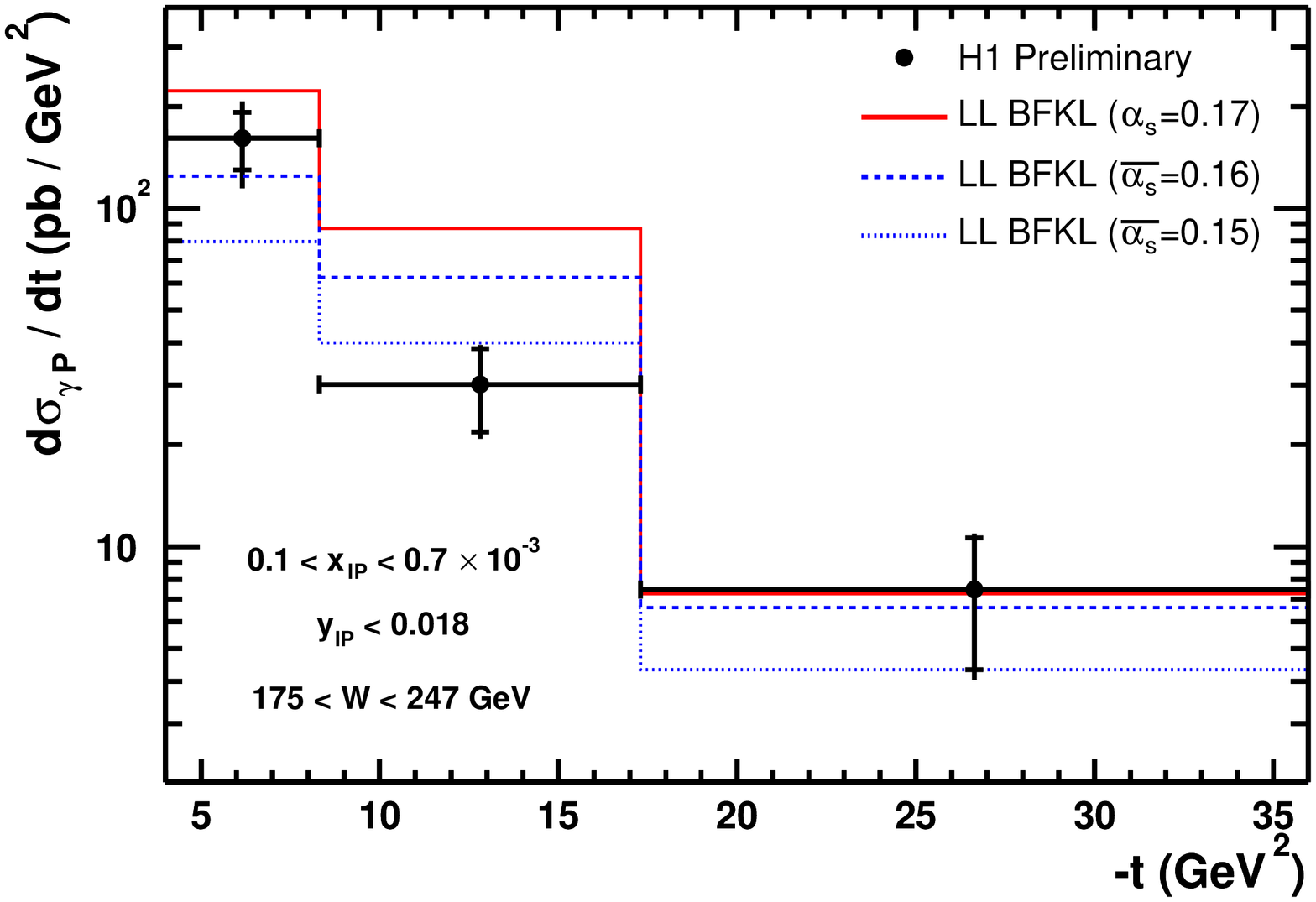,angle=270, width=.48\textwidth}
\caption{The $\gamma p$ cross section differential in $\xpom$ (upper)
 and in $t$ (lower).  The inner
 error bars show the statistical error and the outer error bars show the
 statistical and  systematic errors added in quadrature. The dotted line
 shows the LL BFKL prediction, for different choices of
 $\overline{\alpha_s}$.
}
\label{fig:hight}
\end{figure}

The $\gamma p$ cross section differential in $\xpom$, in the range
$175 < W < 247$~GeV, $p_{t(\gamma)} > 2$ GeV, $\ypom < 0.018$, is
shown in Fig.~\ref{fig:hight} (upper). 
The largest systematic error is due to the uncertainty on the
noise subtracted from the LAr calorimeter.  The total systematic
uncertainty is small compared to the statistical error. Also shown are
the Monte Carlo prediction based on 
the leading logarithmic approximation (LL) of BFKL~\cite{Cox:1999kv}. 
The data are sensitive to the choice of the
$\overline{\alpha_s}$ parameter in the prediction through the slope in 
$\xpom$ related to the BFKL pomeron intercept and the normalisation.
A choice of $\overline{\alpha_s} = 0.17$ gives a reasonable 
description of the data in both normalisation and shape. This is in
comparison to the
value $\overline{\alpha_s} = 0.18$ as used for recent H1 measurements
on $J/\Psi$ production~\cite{Aktas:2003zi,Gaps}.
 
\noindent
The $\gamma p$ cross section differential in the squared 4 momentum
transfer $t$ between proton and the incoming photon (where in
photoproduction $-t \sim p_{t(\gamma)}^2$), in the range $175 < W <
247$~GeV, $0.0001 < \xpom < 0.0007$, $\ypom < 0.018$, is shown in
Fig. \ref{fig:hight} (lower). In this case,
the agreement in the shape of the cross section between the
Monte Carlo predictions and the data is perhaps more questionable, for
all values of $\overline{\alpha_s}$ chosen here.

It is worth bearing in mind that there may be important contributions
from higher order effects beyond the LL, so strong statements about
the agreement may be premature. However, the
fact that the measured cross section exhibits such a dramatic rise
with energy ($W\sim 1/\xpom$) is a striking result, 
and there is an reasonable overall
agreement with the LL BFKL predictions.

\section{Deeply Virtual Compton Scattering}
\label{sec:dvcs}

First cross section measurements of the DVCS process were published 
by H1~\cite{h1-dvcs} and ZEUS~\cite{zeus-dvcs}. 
Here, the new H1 measurement is reported, in an extended kinematic
range: $4<Q^2<80$ GeV$^2$, $30<W<140$ GeV and $|t|<1$ GeV$^2$, using 
integrated luminosity of 26 $\pbinv$ of data taken during the year 2000 
(i.e. 3.5 times larger than the previously published by
H1~\cite{h1-dvcs}).

At these small values of $t$ the reaction
$e p \rightarrow e \gamma p\,$
is dominated by the purely electromagnetic
Bethe-Heitler (BH) process whose cross section,
depending only on QED calculations and proton elastic form factors, is
precisely known and therefore can be subtracted.
To enhance the ratio of selected DVCS
events to BH events the outgoing photon is selected in the
forward, or outgoing proton, region with transverse momentum larger than 
$2\,{\rm GeV}$. Large values of
the incoming photon virtuality $Q^{2}$ are selected by detecting the
scattered
electron in the SPACAL calorimeter with energy larger than $15\,{\rm
GeV}$. The outgoing proton escapes down the
beam-pipe in the forward direction.

\begin{figure}[htbp]
 \begin{center}
  \epsfig{file=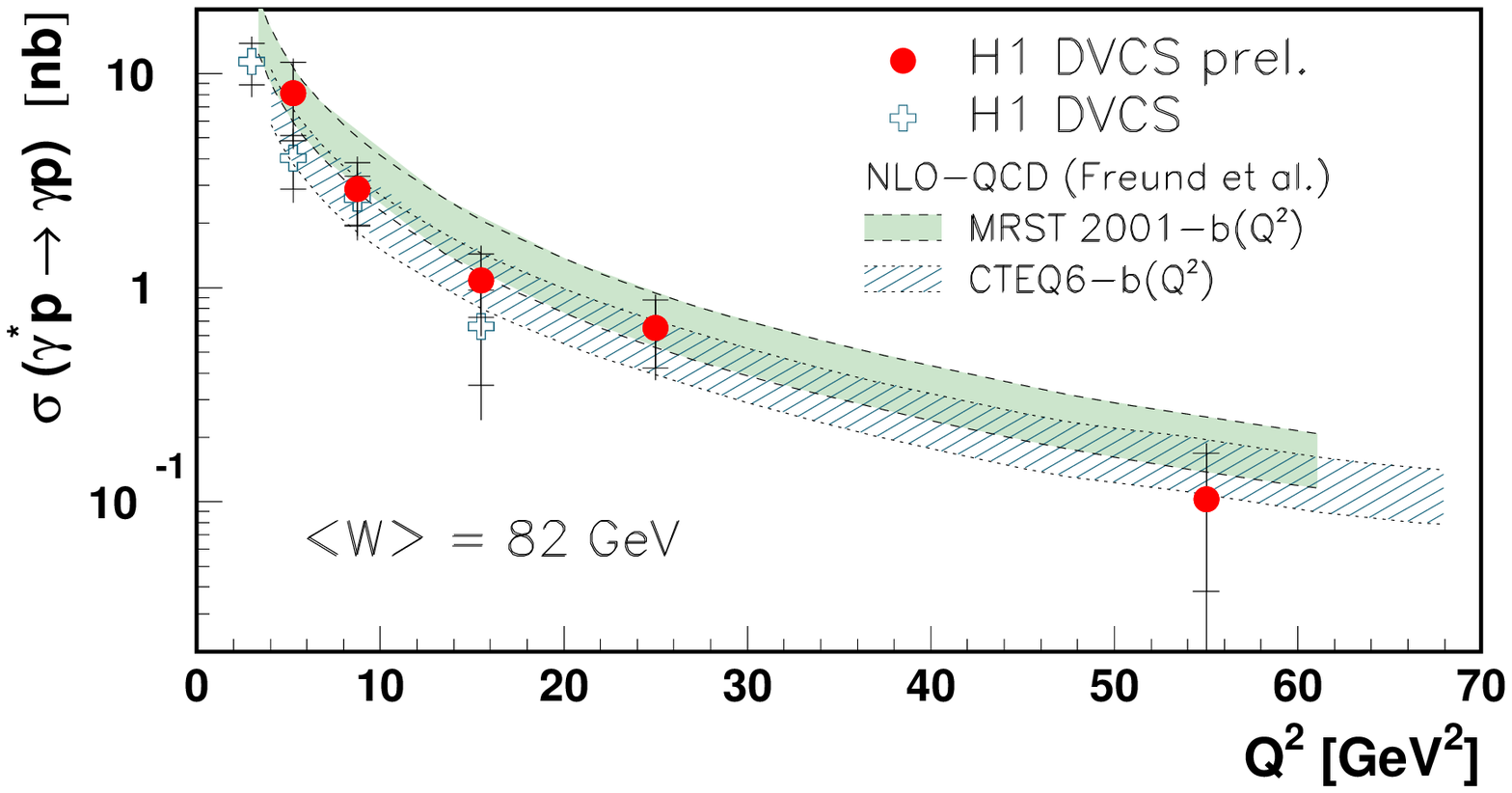,width=0.49\textwidth}
  \epsfig{file=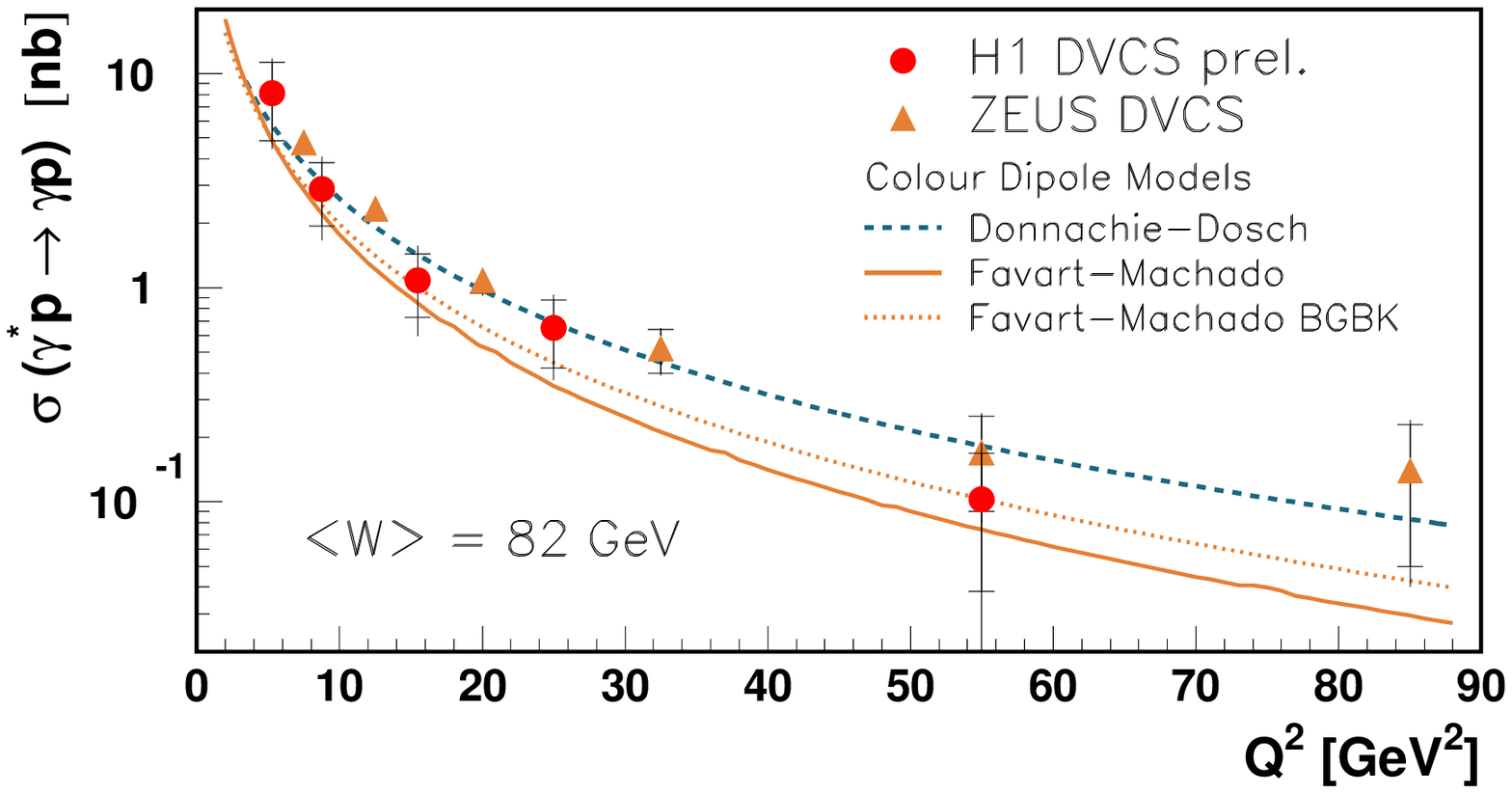,width=0.49\textwidth}
 \end{center}
 \caption{$\gamma^* p \rightarrow \gamma p$
 cross section as a function of $Q^2$ for $<W>=82$~GeV.
 The inner error bars are statistical and
 the full error bars include the systematic errors added in quadrature.
 The measurement is compared (upper plot)
 to the NLO QCD prediction~\cite{Freund:2001hm,Freund:2001hd} 
 using 
 GPD parametrisations based on MRST2001 and CTEQ6~\cite{Freund:2002qf}
 The bands correspond to $b_0$ values between 5 and 9~GeV$^{-2}$.
 On the lower plot, the measurement is compared to 
 two different Colour Dipole models predictions, by Donnachie and
 Dosch~\cite{Donnachie:2000px} and by Favart, Machado~\cite{Favart:2003cu}
 at the fixed value $b=7$ GeV$^{-2}$. The BGBK notation indicates
 the additional DGLAP evolution of the dipole cross section 
 added to the basic prediction. 
}
\label{fig:gpsigq2}
\end{figure}

\begin{figure}[tbh]
 \begin{center}
  \epsfig{file=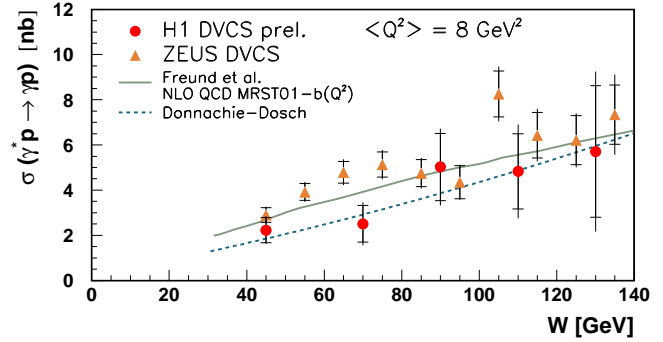,width=0.49\textwidth}
 \end{center}
 \caption{$\gamma^* p \rightarrow \gamma p$
cross section as a function of $W$ for $<Q^2>=8$~GeV$^2$.
 The measurement is compared
 to the NLO QCD prediction~\cite{Freund:2001hm,Freund:2001hd}
 using GPD parametrisation based on MRST2001~\cite{Freund:2002qf}
 and to the Colour Dipole models prediction of Donnachie and
 Dosch~\cite{Donnachie:2000px}.
}
\label{fig:gpsigw}
\end{figure}

The $\gamma^* p$ cross section for the DVCS process is shown in
Fig.~\ref{fig:gpsigq2} as a function
of $Q^2$ for $W=82$~GeV, 
and in Fig.~\ref{fig:gpsigw} as a function of $W$ for $Q^2=8$~GeV$^2$.
In the upper plot of Fig.~\ref{fig:gpsigq2} the measurement is compared
to the NLO QCD prediction~\cite{Freund:2001hm,Freund:2001hd} using two different
GPD parametrisations~\cite{Freund:2002qf}.
The  $t$ dependence is parametrised as $e^{-b|t|}$, with $b=b_0(1-0.15
\log(Q^2/2))$~GeV$^{-2}$. 
The classic PDF $q(x,\mu^2)$ of MRST2001 and CTEQ6 are used in the DGLAP region
($x>\xi$) such that $\cal H$, which is the only important GPD at small $x$ 
is given at the scale $\mu$ by:
${\cal H}^q(x,\xi,t;\mu^2)=q(x;\mu^2) \, e^{-b|t|}$ for quark singlet and
$ {\cal H}^g(x,\xi,t;\mu^2)=x\ g(x;\mu^2) \, e^{-b|t|}$ for gluons,
i.e. independent of the skewing parameter $\xi$.
Keeping this parametrisation in the ERBL region ($|x|<\xi$) would lead 
to a prediction overshooting the data by a factor 4-5.
Therefore, a parametrisation is proposed by the
authors to suppress the region of very small $x$ (for details
see~\cite{Freund:2002qf}). This emphasises the interesting sensitivity
to the ERBL region.
The NLO QCD predictions are in good
agreement with the data, for both GPD parametrisations. Since the main
difference between the two parametrisations resulting in the
normalisation, it emphasizes the need for a direct $t$ dependence
measurement. 

In the lower plots of Fig.~\ref{fig:gpsigq2} the measurement is compared
to two different Colour Dipole models predictions, by Donnachie and
Dosch~\cite{Donnachie:2000px} and by Favart and Machado~\cite{Favart:2003cu}.
They are based on a factorisation into the incoming photon
wave function, a $q\bar q$-p cross section and the outgoing photon wave
function. The models differ in the way the quark
dipole cross section is parametrised. Donnachie and
Dosch~\cite{Donnachie:2000px}
basically connect a soft Pomeron with large dipole size and a hard
Pomeron with
small dipole size. Favart and Machado~\cite{Favart:2003cu} apply the
saturation model
of Golec-Biernat et al.~\cite{Golec-Biernat:1999qd} to the DVCS process,
with a possible DGLAP evolution~\cite{Bartels:2002es} quoted BGBK on
the plot.
In both cases an exponential $t$-dependence, $e^{-b|t|}$, is
assumed.
All presented Colour Dipole model predictions describe well the data in
shape and in normalisation for the same value of $b=7$ GeV$^{-2}$.

The new measurement is also compared to the previous
measurement by H1~\cite{h1-dvcs} 
and to the ZEUS measurement~\cite{zeus-dvcs}.
The two H1 measurements are in good agreement. The new H1 measurement
is in fair agreement with ZEUS results except for $W\sim70$ GeV, where
H1 points are lower by about two standard deviations.


\end{document}